\documentclass[a4paper,11pt]{article}

\usepackage[utf8]{inputenc}
\usepackage[T1]{fontenc}
\usepackage[english]{babel}

\usepackage{indentfirst}
\usepackage{amsmath}
\usepackage{amssymb}
\usepackage{amsthm}
\usepackage{graphicx}
\usepackage{subfigure}
\usepackage{psfrag}{}
\usepackage{color}
\usepackage{pgf}

\allowhyphens

\newcommand{\eps}{\varepsilon}

\newcommand{\ordo}{\mathcal{O}}

\newcommand{\ket}[1]{{\left|#1\right\rangle}}
\newcommand{\bra}[1]{{\left\langle #1\right|}}

\newcommand{\skalarszorzat}[2]{{\langle #1 | #2 \rangle}}


\makeatother

\begin{document}

\author{Bal\'azs Pozsgay$^1$\\
~\\
 $^{1}$MTA-BME \textquotedbl{}Momentum\textquotedbl{} Statistical
Field Theory Research Group\\
1111 Budapest, Budafoki \'ut 8, Hungary
}
\title{Overlaps between eigenstates of the XXZ spin-1/2 chain and a class of
simple product states}

\maketitle

\abstract{We consider a class of quantum quenches in the spin-$1/2$
  XXZ chain, where the initial state is of a simple product form.
 Specific examples are the N\'eel state, the dimer state and
  the $q$-deformed dimer state.
We compute determinant formulas for finite volume overlaps
  between the initial state and arbitrary eigenstates of the spin
  chain Hamiltonian. These results could serve as a basis for
  calculating the time dependence of correlation functions following the
  quantum quench.
}

\numberwithin{equation}{section}

\section{Introduction}

One dimensional exactly solvable quantum mechanical models and in particular the integrable
spin chains attracted a lot of interest since Bethe's famous solution of
the spin-$1/2$ XXX Heisenberg model in 1931 \cite{XXX}. A common property of
these theories is that the spectrum of the Hamiltonian can be computed
with exact methods \cite{sutherland-book} and even thermodynamic
quantities can be obtained in the infinite volume limit
\cite{takahashi-book,kluemper-review}. While it was believed for a long time that
the Bethe Ansatz is not effective in calculating correlation functions,
this situation drastically changed over the last 25 years. 
The discovery of
determinant formulas for certain overlaps between Bethe
states \cite{slavnov-overlaps} and the solution of the so-called
inverse problem of the Algebraic Bethe Ansatz
\cite{maillet-inverse-scatt,goehmann-korepin-inverse} lead
to multiple integral formulas for short-range correlations in the
ground state of the anti-ferromagnetic XXZ model
\cite{KitanineMailletTerras-XXZ-corr1,spin-spin-XXZ}, which were later
extended to the finite temperature case using the so-called Quantum
Transfer Matrix method \cite{QTM1}. Subsequently it was discovered
that the multiple integral formulas can be factorized: it is possible
to express them as sums of products of simple integrals (see
\cite{XXZ-factorization-recent-osszefoglalo} and references therein).

This tremendous progress 
concerns correlation functions in thermal equilibrium. 
On the other hand, in recent years non-equilibrium problems such
as quantum quenches and the questions of thermalization stood in the forefront of
research of both integrable and non-integrable models 
\cite{Silva-quench-colloquium}. This motivates the development of
exact methods for correlation functions of integrable models in
non-equilibrium situations. 

While there have been recent advances
on quantum quenches in the 1D Bose-gas
\cite{2013PhRvA..87e3628I,marci-ll-quench1,spyros2,marci-ll-quench2,caux-stb-LL-BEC-quench}
there are practically no exact results available for the interacting XXZ spin
chain. The long-time limit of local observables was considered in the papers
\cite{sajat-xxz-gge,essler-xxz-gge}, but these works bypass the derivation of
the actual time dependence of the correlators and give predictions
based on the so-called Generalized Gibbs Ensemble (GGE) hypothesis
\cite{rigol-gge}. This hypothesis states that in integrable models the
long-time limit of observables follows from a statistical physical
ensemble incorporating all local conserved charges with
appropriate Lagrange-multipliers, which are fixed by the expectation
values of the charges in the initial state. Whereas the GGE was shown
to be valid for models equivalent to free-fermions
\cite{ising-quench-1,ising-quench-2}, there are no actual proofs
available for the interacting case and the predictions of
\cite{sajat-xxz-gge,essler-xxz-gge} have not yet been confirmed by
independent methods.
Recent exact result for quantum quenches in the XXZ chain also include a non-linear integral equation
for the so-called dynamical free energy (the Loschmidt echo at
imaginary times) \cite{sajat-BQTM}. While
this quantity is not directly relevant to the correlation functions,
the paper \cite{sajat-BQTM} showed that the techniques of the Boundary Algebraic Bethe Ansatz
can be used in quench problems. We should note that the dynamical free
energy was considered also in \cite{Fagotti-Loschmidt} with independent methods
based on the GGE hypothesis.

One route towards time dependent observables  after a
quantum quench is through the form factor expansion. Assuming that at
$t=0$ the system is in a state $\ket{\Psi_0}$, and for $t>0$ the time
evolution is governed by a Hamiltonian $H$, inserting two complete sets
of normalized states leads to
\begin{equation}
\label{FF}
\ordo(t)=\sum_{n,m}
  \skalarszorzat{\Psi_0}{n}\bra{n}\ordo\ket{m}\skalarszorzat{m}{\Psi_0}
e^{it(E_n-E_m)}.
\end{equation}
The
 ingredients in the formula above are the energy eigenvalues,
the matrix elements of the operator (the form factors) and
the exact overlaps between the eigenstates and the initial
state. Although the Bethe Ansatz provides the exact spectrum and there are
results available for the form factors too
\cite{maillet-inverse-scatt,KitanineMailletTerras-XXZ-corr1}, the overlaps are typically 
very hard to compute. 
If the state $\ket{\Psi_0}$ is of a sufficiently simple
form, then the scalar product with a Bethe state can be written down
as a sum over permutations, but such formulas are inconvenient for
further analytical or numerical calculations.

In the present work we compute determinant formulas for the overlaps
between Bethe states and a restricted 
class of initial states which are products of simple two-site
vectors. The specific examples are the N\'eel state, the dimer and 
$q$-deformed dimer states (the definitions are given below). Our
results could serve as a basis for the evaluation of the 
expansion \eqref{FF} with either analytical or numerical methods \cite{quench-action}.

The structure of the paper is as follows. In Section 2. we review
the Bethe Ansatz solution of the XXZ chain and provide the necessary
definitions. In Section 3 we compute the overlaps for generic
anisotropy $\Delta$. Section 4 includes numerical results for the
overlaps between the N\'eel state and the ground states in the
anti-ferromagnetic regime $\Delta>-1$. Also we present a comparison to
the exact result of \cite{sajat-BQTM}
for the infinite volume limit of the ground state overlap with the N\'eel.
 Section 5 concerns the XXX limit. Finally we conclude in
Section 6.

\section{The quench problem and the Bethe Ansatz}

Consider the anti-ferromagnetic spin-$1/2$ XXZ Heisenberg model on a chain of length
$2M$ with periodic boundary conditions. The Hamiltonian reads
\begin{equation}
  \label{XXZ-H}
  H_{XXZ}=\sum_{j=1}^{2M}
  (\sigma^x_j\sigma^x_{j+1}+\sigma^y_j\sigma^y_{j+1}+\Delta
(\sigma^z_j\sigma^z_{j+1}-1)).
\end{equation}
Here $\Delta$ is the anisotropy parameter; we will consider both the
massive ($\Delta>1$) and massless ($-1<\Delta\le 1$) regimes.

This Hamiltonian can be diagonalized by the Bethe Ansatz \cite{XXX,XXZ1,XXZ2,XXZ3}. 
As usual we choose the ferromagnetic state $\ket{F_+}$ with all spins up as the
reference state and construct interacting spin waves above this state.
The explicit coordinate space wave function for a state with $M$ down spins is
\begin{equation}
  \label{BA2}
  \tilde\Psi_{2M}(\lambda_1,\dots,\lambda_M|s_1,\dots,s_M)=\sum_{P\in\sigma_M} 
\prod_j  F(\lambda_{P_j},s_j) 
\prod_{j>k} \frac{\sinh(\lambda_{P_j}-\lambda_{P_k}-\eta)}{\sinh(\lambda_{P_j}-\lambda_{P_k})}
\end{equation}
with
\begin{equation}
\label{Fs}
  F(\lambda,s)=\sinh(\eta)
\sinh^{s-1}(\lambda+\eta/2)
\sinh^{2N-s}(\lambda-\eta/2).
\end{equation}
Here $s_j$ denote the positions of the down spins, and we assume
$s_j<s_k$ for $j<k$. The parameter
$\eta$ is given by the relation $\Delta=\cosh\eta$ and the rapidities
$\{\lambda_j\}$ characterize the spin waves. The state \eqref{BA2} is
an eigenstate if the Bethe equations hold:
\begin{equation}
  \label{BAe}
\left(  \frac{\sinh(\lambda_j-\eta/2)}{\sinh(\lambda_j+\eta/2)}\right)^{2M}
\prod_{k\ne j}
\frac{\sinh(\lambda_j-\lambda_k+\eta)}{\sinh(\lambda_j-\lambda_k-\eta)}=1.
\end{equation}
In this case the energy is given by
\begin{equation}
\label{BAee}
  E=\sum_j e(\lambda_j),\quad\text{where}\qquad
e(u)=\frac{4\sinh^2\eta}{\cosh(2u)-\cosh\eta}.
\end{equation}
In the regime $\Delta>1$ the typical one-string solutions to the Bethe
equations lie on the imaginary axis, while for $\Delta<1$ 
 they are on the real axis. The $\Delta=1$ case has to be treated
 separately, see section \ref{sec:xxx}.

Consider a non-equilibrium situation where at $t=0$ the system is
prepared in  a state $\ket{\Psi_0}$ and for $t>0$ it evolves according
to the Hamiltonian \eqref{XXZ-H}. The time evolution of a physical
observable $\ordo$ can be computed by inserting two complete sets of
(not normalized)
states:
\begin{equation}
\label{ot}
\ordo(t)=\sum_{\{\lambda\}_M} \sum_{\{\mu\}_M}  
\frac{\skalarszorzat{\Psi_0}{\{\lambda\}_M}
\bra{\{\lambda\}_M}\ordo\ket{\{\mu\}_M}
\skalarszorzat{\{\mu\}_M}{\Psi_0}}
{\skalarszorzat{\{\lambda\}_M}{\{\lambda\}_M}
\skalarszorzat{\{\mu\}_M}{\{\mu\}_M}
}
 \exp\left(it \sum_j(e(\lambda_j)-e(\mu_j))\right).
\end{equation}
Here we assumed for simplicity that  $\ket{\Psi_0}$ has fixed
magnetization equal to $0$. 
The essential ingredients in the
formula above are the energy eigenvalues, the form factors and the normalized overlaps. While the
Bethe Ansatz is very efficient in computing the energies through
\eqref{BAe}-\eqref{BAee}, and there are determinant formulas available
for form factors too
\cite{maillet-inverse-scatt,KitanineMailletTerras-XXZ-corr1}, there
are practically no results available for the overlaps.

In this work we consider three specific choices for 
$\ket{\Psi_0}$ and compute determinant formulas for the overlap with
the Bethe states. In all 
three cases there exists a local 
Hamiltonian with a gapped spectrum and a twofold degenerate ground
state level, which is spanned by $\ket{\Psi_0}$ and $P\ket{\Psi_0}$
with $P$ being the translation operator by one site. The three vectors
and the corresponding Hamiltonians are:
\begin{itemize}
\item The N\'eel state:
  \begin{equation*}
    \ket{\Psi_0} =\ket{N}\equiv \otimes^{M} \ket{+-},    
  \end{equation*}
which is a ground state of the XXZ Hamiltonian in the
$\Delta\to \infty$ limit.
\item The dimer state:
  \begin{equation*}
      \ket{\Psi_0} =\ket{D}\equiv \otimes^{M}
  \frac{\ket{+-}-\ket{-+}}{\sqrt{2}},
  \end{equation*}
which is a ground state of the Majumdar-Ghosh Hamiltonian \cite{MG}. 
\item The $q$-deformed dimer state:
  \begin{equation*}
      \ket{\Psi_0} =\ket{qD}\equiv \otimes^{M}
  \frac{q^{1/2}\ket{+-}-q^{-1/2}\ket{-+}}{\sqrt{|q|+1/|q|}},
  \end{equation*}
where $q$ is given by the relation $\Delta=(q+1/q)/2$. This vector is a ground
state of the $q$-deformed Majumdar-Ghosh Hamiltonian \cite{q-dimer}.
\end{itemize}

The overlaps are in principle known because the coordinate Bethe
Ansatz provides the exact wave functions.
For example in the case of the N\'eel state we have
\begin{equation}
  \label{BA3}
\skalarszorzat{N}{\{\lambda\}_M}=
\sum_{P\in\sigma_M} 
\prod_j  F(\lambda_{P_j},2j) \prod_{j>k}
\frac{\sinh(\lambda_{P_j}-\lambda_{P_k}-\eta)}{\sinh(\lambda_{P_j}-\lambda_{P_k})}
\end{equation}
with $F(u,s)$ given by \eqref{Fs}. However, such representations are not
convenient for either analytical or numerical treatment, because the
number of terms grows as $M!$ and there are no evident ways to sum
them up.
On the other hand, formulas like \eqref{BA3}
can be used to check the determinant formulas calculated below. We
wish to note that simple relations can be established between the
overlaps with the different initial states based on the coordinate
Bethe Ansatz wave function. For example 
\begin{equation}
  \label{DN}
\skalarszorzat{D}{\{\lambda\}_M}=
\skalarszorzat{N}{\{\lambda\}_M}\times
\prod_{j=1}^M \frac{1}{\sqrt{2}}\left(
1-\frac{\sinh(\lambda_j-\eta/2)}{\sinh(\lambda_j+\eta/2)}
\right).
\end{equation}
Such relations can also be used as a check of the formulas presented below.

In the next section we will make use of the Algebraic Bethe Ansatz
which is an alternative method for the diagonalization of the XXZ
Hamiltonian. Its basic object is 
the monodromy matrix which is defined as 
\begin{equation*}
  T(u)=L_{1}(u)\dots L_{2M}(u),
\end{equation*}
where $L_j(u)$ are local Lax-operators given by
\begin{equation}
\label{LL}
  L_j(u)=R_{0j}(u-\eta/2),
\end{equation}
and the $R$-matrix is
\begin{equation}
  R(u)=
  \begin{pmatrix}
    \sinh(u+\eta) & & &\\
& \sinh(u)  & \sinh(\eta) & \\
& \sinh(\eta) & \sinh(u) & \\
& & & \sinh(u+\eta)
  \end{pmatrix}.
\label{R}
\end{equation}
In \eqref{LL} the index $j$ refers to the spin on site $j$ whereas $0$
refers to the auxiliary space. The monodromy matrix is written in
auxiliary space as
\begin{equation}
\label{TTT}
  T(u)=
  \begin{pmatrix}
    A(u) & B(u) \\
C(u) & D(u)
  \end{pmatrix}.
\end{equation}
Eigenstates with a total number of $M$ down spins are then constructed
as
\begin{equation*}
  \ket{\{\lambda\}_M}=\prod_{j=1}^M B(\lambda_j) \ket{F_+}.
\end{equation*}
It can be shown that with the present normalizations these states are
exactly identical to those given by the
 expression \eqref{BA2}.

Besides the overlaps it is also essential to know the norms of Bethe
states \cite{gaudin-LL-norms,korepin-norms}. It was derived in
\cite{korepin-norms} that if the rapidities 
satisfy the Bethe equations then the norm is given by the determinant formula
\begin{equation}
\label{norm}
\begin{split}
&  \skalarszorzat{\lambda_1,\dots,\lambda_M}{\lambda_1,\dots,\lambda_M}=\\
&\hspace{0.5cm}=\sinh^{M}(\eta)\prod_{j}
(\sinh(\lambda_j+\eta/2)\sinh(\lambda_j-\eta/2))^{2M}
\times\prod_{j\ne k} f(\lambda_j,\lambda_k)\times \det G,
\end{split}
\end{equation}
where 
\begin{equation*}
  f(u)=\frac{\sinh(u+\eta)}{\sinh(u)},
\end{equation*}
and $G$ is the Gaudin matrix:
\begin{equation*}
  G_{jk}=\delta_{j,k}\left(2M \varphi(\lambda_j,\eta/2)-\sum_l
    \varphi(\lambda_j-\lambda_l,\eta)\right)+
\varphi(\lambda_j-\lambda_k,\eta)
\end{equation*}
with
\begin{equation*}
  \varphi(a,b)= \frac{\sinh(2b)}{\sinh(a-b)\sinh(a+b)}.
\end{equation*}

\section{Boundary Algebraic Bethe Ansatz for the overlaps}

\label{sec:BABA}

In this section we derive determinant formulas for the overlaps
between arbitrary Bethe states and the three initial states specified
above.  The method we apply is the Boundary Algebraic Bethe
Ansatz, which was originally devised to diagonalize open spin chains with
boundary magnetic fields and to compute correlation functions in these
models  \cite{sklyanin-boundary,openXXZ1}.
For a thorough explanation of the Boundary Algebraic Bethe Ansatz we
refer the reader to \cite{openXXZ1}; here we only use the relevant
results of \cite{openXXZ1}.

Consider the boundary transfer matrix of a spin chain
of length $M$:
\begin{equation}
\label{boundaryT}
  \mathcal{R}(u)=\text{Tr}_0\left\{ K^+(u) T_1(u)K^-(u)  T_2(u)\right\}.
\end{equation}
Here
\begin{equation}
\label{T1}
  T_1(u)=\tilde L_M(u)\dots \tilde L_1(u),
\end{equation}
where $\tilde L_j(u)$ are given by
\begin{equation}
\label{Lis}
  \tilde L_j(u)=R_{0j}(u-\xi_j)
\end{equation}
and 
\begin{equation*}
  T_2(u)=\gamma(u) \sigma_0^y T^{t_0}_1(-u) \sigma_0^y.
\end{equation*}
The parameters $\xi_j$ are inhomogeneities which will be specified
below. The function $\gamma(u)$ is related to the crossing properties
of the $R$-matrix; in the present normalization  $\gamma(u)=(-1)^M$. For
simplicity we assume that $M$ is even and we will use the relation
\begin{equation*}
  (-1)^M=1
\end{equation*}
throughout this work. The operator in \eqref{T1} is depicted in figure \ref{fig:mono}.

The matrices $K^\pm$ in \eqref{boundaryT} are the diagonal solutions to the reflection
equation \cite{sklyanin-boundary,openXXZ1}:
\begin{equation*}
  K^\pm(u)=K(u\pm \eta/2,\xi_\pm)\quad\text{with}\quad
 K(u,\xi)=
 \begin{pmatrix}
   \sinh(\xi+u) & 0 \\
0 & \sinh(\xi-u)
 \end{pmatrix}
.
\end{equation*}
In the original problem of the boundary spin chain the parameters
$\xi_\pm$ are related to the boundary magnetic fields. Here
we leave them unspecified for the moment.

The common eigenstates of the operators $\mathcal{R}(u)$ can be created
from the ferromagnetic reference state $\ket{F_+}=\ket{++\dots}$ as
\begin{equation}
\label{bABAstates2}
  \ket{\{\lambda\}_n}=\prod_{j=1}^n \mathcal{B}_+(\lambda_j) \ket{F_+},
\end{equation}
where the $\mathcal{B}_+(u)$ operators are defined through
\begin{equation*}
T^{t_0}_1(\lambda)K_+^{t_o}(\lambda) T_2^{t_0}(\lambda)
=
  \begin{pmatrix}
   \mathcal{A}_+(u) & \mathcal{C}_+(u) \\
\mathcal{B}_+(u) & \mathcal{D}_+(u)
  \end{pmatrix}.
\end{equation*}
Writing out the components we obtain
\begin{equation}
  \label{Bmin2}
   \mathcal{B}_+(u)=B_1(u)D_1(-u) k_1(u) -D_1(u)B_1(-u)k_2(u)
\end{equation}
with
\begin{equation}
\label{k1k2}
  k_1(u)=\sinh(\xi_++u+\eta/2) \qquad
  k_2(u)=\sinh(\xi_+-u-\eta/2).
\end{equation}
Alternatively, eigenstates could be created from the spin-reversed
reference state $\ket{F_-}=\ket{--\dots}$ by the action of the operators
$\mathcal{C}_+(u)$ for which we have
\begin{equation}
  \label{Bmin3}
   \mathcal{C}_+(u)=-A_1(u)C_1(-u) k_1(u) +C_1(u)A_1(-u)k_2(u).
\end{equation}

\begin{figure}
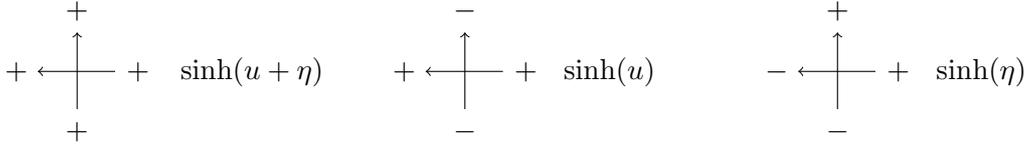

\centering
\begin{pgfpicture}{0cm}{-0.2cm}{14cm}{2.2cm}
\pgfsetendarrow{\pgfarrowto}

\pgfline{\pgfxy(1.5,1)}{\pgfxy(0.5,1)}  
\pgfline{\pgfxy(1,0.5)}{\pgfxy(1,1.5)}

\pgfputat{\pgfxy(1.8,1)}{\pgfbox[center,center]{$+$}}
\pgfputat{\pgfxy(1,0.2)}{\pgfbox[center,center]{$+$}}
\pgfputat{\pgfxy(1,1.8)}{\pgfbox[center,center]{$+$}}
\pgfputat{\pgfxy(0.2,1)}{\pgfbox[center,center]{$+$}}

\pgfputat{\pgfxy(3.3,1)}{\pgfbox[center,center]{$\sinh(u+\eta)$}}

\pgfline{\pgfxy(6.6,1)}{\pgfxy(5.6,1)}  
\pgfline{\pgfxy(6.1,0.5)}{\pgfxy(6.1,1.5)}

\pgfputat{\pgfxy(6.9,1)}{\pgfbox[center,center]{$+$}}
\pgfputat{\pgfxy(6.1,0.2)}{\pgfbox[center,center]{$-$}}
\pgfputat{\pgfxy(6.1,1.8)}{\pgfbox[center,center]{$-$}}
\pgfputat{\pgfxy(5.3,1)}{\pgfbox[center,center]{$+$}}

\pgfputat{\pgfxy(7.4,1)}{\pgfbox[left,center]{$\sinh(u)$}}

\pgfline{\pgfxy(11.5,1)}{\pgfxy(10.5,1)}  
\pgfline{\pgfxy(11,0.5)}{\pgfxy(11,1.5)}

\pgfputat{\pgfxy(11.8,1)}{\pgfbox[center,center]{$+$}}
\pgfputat{\pgfxy(11,0.2)}{\pgfbox[center,center]{$-$}}
\pgfputat{\pgfxy(11,1.8)}{\pgfbox[center,center]{$+$}}
\pgfputat{\pgfxy(10.2,1)}{\pgfbox[center,center]{$-$}}

\pgfputat{\pgfxy(12.3,1)}{\pgfbox[left,center]{$\sinh(\eta)$}}
\end{pgfpicture}
\caption{The 6-vertex model weights as given by the trigonometric $R$-matrix \eqref{R}. Here $u$ is attached to the horizontal line and $0$ to the vertical
one.}
\label{fig:6}
\end{figure}

\begin{figure}
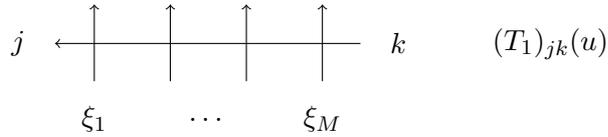

\centering
\begin{pgfpicture}{4cm}{-0.2cm}{11cm}{2.2cm}
\pgfsetendarrow{\pgfarrowto}

\pgfline{\pgfxy(8.5,1)}{\pgfxy(4.5,1)}  
\pgfline{\pgfxy(5,0.5)}{\pgfxy(5,1.5)}
\pgfline{\pgfxy(6,0.5)}{\pgfxy(6,1.5)}
\pgfline{\pgfxy(7,0.5)}{\pgfxy(7,1.5)}
\pgfline{\pgfxy(8,0.5)}{\pgfxy(8,1.5)}

\pgfputat{\pgfxy(4,1)}{\pgfbox[center,center]{$j$}}
\pgfputat{\pgfxy(9,1)}{\pgfbox[center,center]{$k$}}
\pgfputat{\pgfxy(11,1)}{\pgfbox[center,center]{$(T_{1})_{jk}(u)$}}

\pgfputat{\pgfxy(5,0)}{\pgfbox[center,center]{$\xi_1$}}
\pgfputat{\pgfxy(6.5,0)}{\pgfbox[center,center]{$\dots$}}
\pgfputat{\pgfxy(8,0)}{\pgfbox[center,center]{$\xi_M$}}

\end{pgfpicture}
\caption{Pictorial depiction of the one-row monodromy matrix which
  enters the definition of the boundary transfer matrix
  \eqref{boundaryT}. The rapidity $u$ is attached to the horizontal
  line, whereas the vertical lines carry inhomogeneities $\xi_j$.
The vertex weights are specified by \eqref{Lis} and the 6-vertex model
weights depicted in fig. \ref{fig:6}.}
\label{fig:mono}
\end{figure}

\begin{figure}
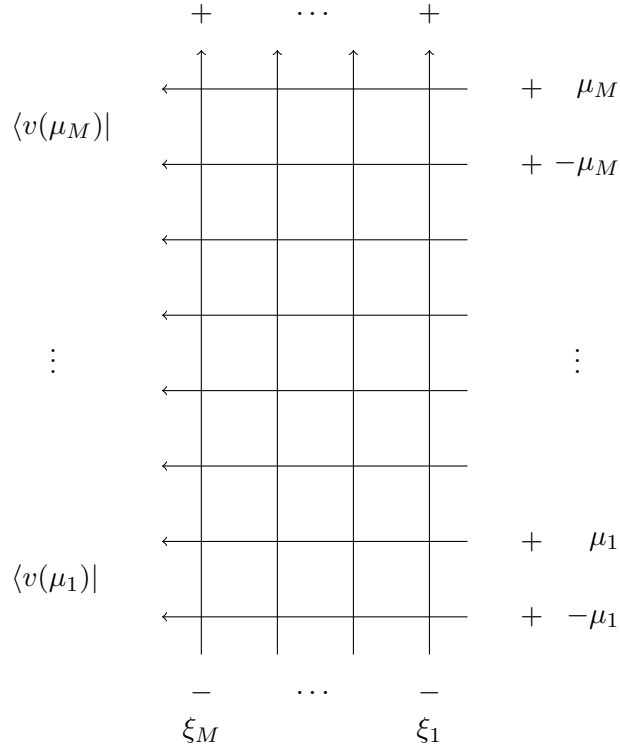

\centering
\begin{pgfpicture}{3cm}{3cm}{12cm}{13cm}

\pgfsetendarrow{\pgfarrowto}

\pgfline{\pgfxy(9.5,5)}{\pgfxy(5.5,5)}
\pgfline{\pgfxy(9.5,6)}{\pgfxy(5.5,6)}
\pgfline{\pgfxy(9.5,7)}{\pgfxy(5.5,7)}
\pgfline{\pgfxy(9.5,8)}{\pgfxy(5.5,8)}
\pgfline{\pgfxy(9.5,9)}{\pgfxy(5.5,9)}
\pgfline{\pgfxy(9.5,10)}{\pgfxy(5.5,10)}
\pgfline{\pgfxy(9.5,11)}{\pgfxy(5.5,11)}
\pgfline{\pgfxy(9.5,12)}{\pgfxy(5.5,12)}

\pgfclearstartarrow

\pgfline{\pgfxy(6,4.5)}{\pgfxy(6,12.5)}
\pgfline{\pgfxy(7,4.5)}{\pgfxy(7,12.5)}
\pgfline{\pgfxy(8,4.5)}{\pgfxy(8,12.5)}
\pgfline{\pgfxy(9,4.5)}{\pgfxy(9,12.5)}

\pgfputat{\pgfxy(9,4)}{\pgfbox[center,center]{$-$}}
\pgfputat{\pgfxy(7.5,4)}{\pgfbox[center,center]{$\dots$}}
\pgfputat{\pgfxy(6,4)}{\pgfbox[center,center]{$-$}}
\pgfputat{\pgfxy(9,3.5)}{\pgfbox[center,center]{$\xi_1$}}
\pgfputat{\pgfxy(6,3.5)}{\pgfbox[center,center]{$\xi_M$}}

\pgfputat{\pgfxy(9,13)}{\pgfbox[center,center]{$+$}}
\pgfputat{\pgfxy(7.5,13)}{\pgfbox[center,center]{$\dots$}}
\pgfputat{\pgfxy(6,13)}{\pgfbox[center,center]{$+$}}

\pgfclearendarrow

\pgfputat{\pgfxy(3.5,5.5)}{\pgfbox[left,center]{$ \bra{v(\mu_1)}$}}
\pgfputat{\pgfxy(4,8.5)}{\pgfbox[left,center]{$ \vdots$}}
\pgfputat{\pgfxy(3.5,11.5)}{\pgfbox[left,center]{$ \bra{v(\mu_M)}$}}

\pgfputat{\pgfxy(10.5,5)}{\pgfbox[right,center]{$+$}}
\pgfputat{\pgfxy(10.5,6)}{\pgfbox[right,center]{$+$}}
\pgfputat{\pgfxy(10.5,11)}{\pgfbox[right,center]{$+$}}
\pgfputat{\pgfxy(10.5,12)}{\pgfbox[right,center]{$+$}}

\pgfputat{\pgfxy(11.5,5)}{\pgfbox[right,center]{$-\mu_1$}}
\pgfputat{\pgfxy(11.5,6)}{\pgfbox[right,center]{$\mu_1$}}
\pgfputat{\pgfxy(11,8.5)}{\pgfbox[right,center]{$\vdots$}}
\pgfputat{\pgfxy(11.5,11)}{\pgfbox[right,center]{$-\mu_M$}}
\pgfputat{\pgfxy(11.5,12)}{\pgfbox[right,center]{$\mu_M$}}

\end{pgfpicture}
\caption{The 6-vertex model configuration whose partition function
  gives the scalar product defined in \eqref{ZZ}. 
}
\label{alap-overlap}
\end{figure}

Consider the scalar product
\begin{equation}
\label{ZZ}
  \mathcal{Z}(\{\mu\}_M,\{\xi\}_M,\xi_+)=\bra{F_+}\mathcal{C}_+(\mu_1)\dots
  \mathcal{C}_+(\mu_M)\ket{F_-}.
\end{equation}
Using  \eqref{Bmin3} 
it can be interpreted as a partition function of a 6-vertex model
with boundary conditions and inhomogeneities defined in figure
\ref{alap-overlap}. The two-site states $\ket{v(u)}$ which
serve as the boundary condition on the left hand side follow from \eqref{k1k2}-\eqref{Bmin3}:
\begin{equation}
  \bra{v(u)}=-\sinh(\xi_++u+\eta/2)\bra{+-}+
\sinh(\xi_+-u-\eta/2)\bra{-+}.
\end{equation}

\begin{figure}
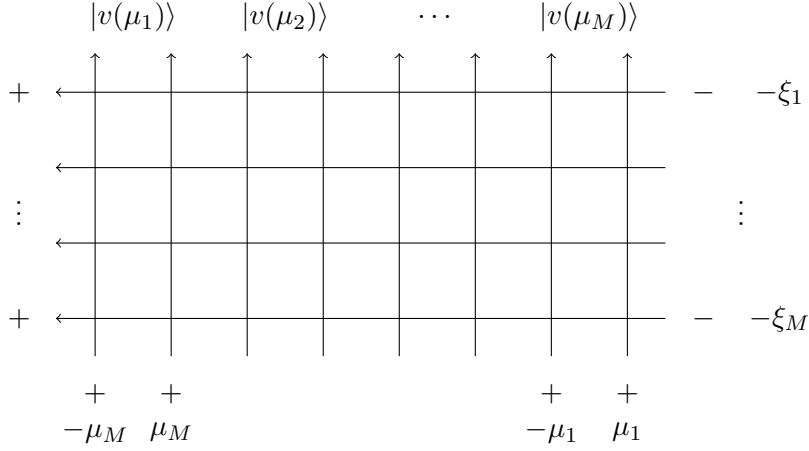

\centering
\begin{pgfpicture}{0cm}{3cm}{12cm}{9cm}

\pgfsetstartarrow{\pgfarrowto}

\pgfline{\pgfxy(1.5,5)}{\pgfxy(9.5,5)}
\pgfline{\pgfxy(1.5,6)}{\pgfxy(9.5,6)}
\pgfline{\pgfxy(1.5,7)}{\pgfxy(9.5,7)}
\pgfline{\pgfxy(1.5,8)}{\pgfxy(9.5,8)}

\pgfputat{\pgfxy(11,8)}{\pgfbox[center,center]{$-\xi_1$}}
\pgfputat{\pgfxy(11,5)}{\pgfbox[center,center]{$-\xi_M$}}
\pgfputat{\pgfxy(10,8)}{\pgfbox[center,center]{$-$}}
\pgfputat{\pgfxy(10,5)}{\pgfbox[center,center]{$-$}}
\pgfputat{\pgfxy(10.5,6.5)}{\pgfbox[center,center]{$\vdots$}}

\pgfputat{\pgfxy(1,8)}{\pgfbox[center,center]{$+$}}
\pgfputat{\pgfxy(1,5)}{\pgfbox[center,center]{$+$}}
\pgfputat{\pgfxy(1,6.5)}{\pgfbox[center,center]{$\vdots$}}

\pgfline{\pgfxy(2,8.5)}{\pgfxy(2,4.5)}
\pgfline{\pgfxy(3,8.5)}{\pgfxy(3,4.5)}
\pgfline{\pgfxy(4,8.5)}{\pgfxy(4,4.5)}
\pgfline{\pgfxy(5,8.5)}{\pgfxy(5,4.5)}
\pgfline{\pgfxy(6,8.5)}{\pgfxy(6,4.5)}
\pgfline{\pgfxy(7,8.5)}{\pgfxy(7,4.5)}
\pgfline{\pgfxy(8,8.5)}{\pgfxy(8,4.5)}
\pgfline{\pgfxy(9,8.5)}{\pgfxy(9,4.5)}

\pgfputat{\pgfxy(2.5,9)}{\pgfbox[center,center]{$\ket{v(\mu_1)}$}}
\pgfputat{\pgfxy(4.5,9)}{\pgfbox[center,center]{$\ket{v(\mu_2)}$}}
\pgfputat{\pgfxy(6.5,9)}{\pgfbox[center,center]{$\dots$}}
\pgfputat{\pgfxy(8.5,9)}{\pgfbox[center,center]{$\ket{v(\mu_M)}$}}
\pgfputat{\pgfxy(2,4)}{\pgfbox[center,center]{$+$}}
\pgfputat{\pgfxy(3,4)}{\pgfbox[center,center]{$+$}}
\pgfputat{\pgfxy(8,4)}{\pgfbox[center,center]{$+$}}
\pgfputat{\pgfxy(9,4)}{\pgfbox[center,center]{$+$}}

\pgfputat{\pgfxy(2,3.5)}{\pgfbox[center,center]{$-\mu_M$}}
\pgfputat{\pgfxy(3,3.5)}{\pgfbox[center,center]{$\mu_M$}}
\pgfputat{\pgfxy(8,3.5)}{\pgfbox[center,center]{$-\mu_1$}}
\pgfputat{\pgfxy(9,3.5)}{\pgfbox[center,center]{$\mu_1$}}

\pgfclearendarrow

\end{pgfpicture}
\caption{
The 6-vertex model configuration obtained from the one in fig. \ref{alap-overlap}
after a reflection along the NW diagonal. Here the horizontal lines
can be interpreted as $B$-operators acting on the ferromagnetic
reference state of an inhomogeneous spin chain. The two-site states
$\ket{v(\mu_j)}$ create a boundary condition at the top, which
in the $\mu_j\to 0$ limit
generates the desired overlaps between Bethe states and the initial
states $\ket{\Psi_0}$.
}
\label{atirva}
\end{figure}

The six-vertex model vertex weights are invariant under the reflection
along the NW diagonal and a simultaneous sign change of the rapidities. 
Performing these operations
we obtain the partition function represented in
fig. \ref{atirva}. Here the horizontal lines can be interpreted as the
$B$-operators of a spin chain of length $2M$ with alternating inhomogeneities given
by $\{\mu_j,-\mu_j\}$.
In the homogeneous limit $\mu_j\to 0$ we have
\begin{equation}
\label{egykoztes}
\lim_{\mu_j\to 0}  \mathcal{Z}(\{\mu\}_M,\{\xi\}_M,\xi_+)=
\bra{\Psi(\xi_+)}B(-\xi_1+\eta/2)B(-\xi_2+\eta/2)\dots B(-\xi_M+\eta/2)\ket{F_+},
\end{equation}
where
\begin{equation*}
\begin{split}
\bra{  \Psi(\xi_+)}&=\otimes^M \bra{v} \\
\bra{v}&=-\sinh(\xi_++\eta/2)\bra{+-}+
\sinh(\xi_+-\eta/2)\bra{-+},
\end{split}
\end{equation*}
and the $B$-operators are defined by the monodromy matrix of the original periodic
problem \eqref{TTT}. The shift of $\eta/2$ in the rapidities in
\eqref{egykoztes} follows from the definition \eqref{LL}.

In the construction above the parameter $\xi_+$ can be chosen
arbitrarily. The N\'eel, the dimer and the $q$-dimer states are
obtained as
\begin{equation}
\label{naigylesz}
  \begin{split}
 \bra{N}&=\frac{\bra{\Psi(\eta/2)}}{\sinh(\eta)^M}\\   
\bra{D}&=\frac{\bra{\Psi(i\pi/2)}}{\cosh(\eta/2)^M}\\
\bra{qD}&=\lim_{\xi\to\infty }
\frac{\bra{\Psi(\xi)}}
{(|\sinh(\xi-\eta/2)|^2+|\sinh(\xi+\eta/2)|^2)^{M/2}}.
  \end{split}
\end{equation}
In the last relation the parameter $q$ is recovered as $q=e^\eta$.

In \cite{tsuchiya-det} Tsuchiya derived a determinant formula for the
partition function \eqref{ZZ}. In the present conventions it reads
\begin{equation}
\label{fo1}
\begin{split}
&\bra{F_+}\mathcal{C}_+(\mu_1)\dots \mathcal{C}_+(\mu_M)\ket{F_-}=
\prod_j  a(\mu_j) a(-\mu_j)\times\\
&\frac{\prod_{j,k} \sinh(\mu_j+\xi_k)\sinh(\mu_j-\xi_k)}
{\prod_{j>k} \sinh(\mu_j-\mu_k)\sinh(\mu_j+\mu_k) \prod_{j<k}\sinh(\xi_j-\xi_k)\sinh(\xi_j+\xi_k-\eta)}
\det I,
\end{split}
\end{equation}
where
\begin{equation}
  I_{jk}=\frac{\sinh\eta \sinh(-2\mu_j-\eta)\sinh(-\xi_++\xi_k-\eta/2)}
{\sinh(\mu_j+\xi_k-\eta)\sinh(\mu_j-\xi_k+\eta)\sinh(-\mu_j+\xi_k)\sinh(\mu_j+\xi_k)},
\end{equation}
and
\begin{equation}
  a(u)=\prod_j \sinh(u-\xi_j+\eta).
\end{equation}
Performing a simple factorization, substituting $\xi_j=-\lambda_j+\eta/2$
 and using the relations \eqref{naigylesz} we find
\begin{equation*}
\begin{split}
 \skalarszorzat{\Psi_0}{\lambda_1,\dots,\lambda_M}=&
\bra{\Psi_0}\prod_{j=1}^M B(\lambda_j)\ket{F_+}=\\
=&\mathcal{P}\mathcal{D}
\frac{\sinh(\eta)^{2M}
\prod_{k}  (\sinh(\lambda_k+\eta/2)\sinh(\lambda_k-\eta/2))^{2M}}
{\prod_{j<k}
\sinh(\lambda_k-\lambda_j)\sinh(\lambda_j+\lambda_k)},
\end{split}
\end{equation*}
where
\begin{equation}
\label{Dlim}
  \mathcal{D}=
\lim_{\mu_j\to 0} \frac{1}{\prod_{j>k}
  \sinh(\mu_j-\mu_k)\sinh(\mu_j+\mu_k)}\det \bar I
\end{equation}
with
\begin{equation*}
\begin{split}
&\bar  I_{jk}=\\ &\hspace{0.2cm}\frac{1}{\sinh(\mu_j+\lambda_k+\eta/2)\sinh(\mu_j-\lambda_k-\eta/2)
\sinh(\mu_j+\lambda_k-\eta/2)\sinh(\mu_j-\lambda_k+\eta/2)},
\end{split}
\end{equation*}
and the factor $\mathcal{P}$ includes all information about the
initial state:
\begin{itemize}
\item For the N\'eel state:
\begin{equation*}
  \mathcal{P}_N=\frac{\prod_k \sinh(\eta/2+\lambda_k)}
{\sinh(\eta)^M }.
\end{equation*}
\item For the dimer state:
  \begin{equation*}
     \mathcal{P}_D=\frac{\prod_k \cosh(\lambda_k)}
{2^{M/2}\cosh(\eta/2)^M }.
  \end{equation*}
\item For the $q$-deformed dimer:
\begin{equation*}
\begin{split}
    \mathcal{P}_{qD}&=\frac{\prod_k e^{\lambda_k}}
{(2\cosh(\eta))^{M/2} }, \quad\text{for}\quad \Delta>1\\
    \mathcal{P}_{qD}&=\frac{\prod_k e^{\lambda_k}}
{2^{M/2} }, \quad\text{for}\quad \Delta<1.\\
\end{split}
\end{equation*}
\end{itemize}

The delicate limit \eqref{Dlim} can be performed in at least two
ways. One possibility is to ``pull out'' a Cauchy determinant from the
matrix $I$ as it was done for a closely related matrix in \cite{sajat-Karol} (see
equation 2.19); the
$\mu\to 0$ limit is trivial afterwards. However, this representation
is plagued by a singularity if $\lambda_j=-\lambda_k$ for some $j,k$,
which is difficult to resolve. Therefore here we take the limit by
explicitly calculating the derivatives of the matrix elements of
$I$.
Using the deformed binomial formula
$\sinh(a+b)\sinh(a-b)=\sinh^2(a)-\sinh^2(b)$ and partial fraction
expansion it follows that
\begin{equation*}
 \bar I_{jk}=
\frac{1}{\sinh(2\lambda_j)\sinh(\eta)}
\left[
\frac{1}{\sinh^2(\lambda_j-\eta/2)-x_k}-
\frac{1}{\sinh^2(\lambda_j+\eta/2)-x_k}
\right] 
\end{equation*}
with $x_k=\sinh^2(\mu_k)$. The $x_j\to x_k$ limit of the determinant
can be performed using the method of
\cite{six-vertex-partition-function} and after some simple
manipulations we arrive at
\begin{equation}
\label{DAN}
  \mathcal{D}=
\prod_k \frac{1}{\sinh\eta\sinh(2\lambda_k)} \times
 \det  L
\end{equation}
where
\begin{equation}
\label{benares4}
\begin{split}
  L_{jk}=q_{2j}(\lambda_k),\quad\text{}\qquad
q_a(u)=\coth^a(u-\eta/2)-\coth^a(u+\eta/2),
\end{split}
\end{equation}  

Our final formula for the overlap with the N\'eel state reads
\begin{equation}
\label{N}
  \skalarszorzat{N}{\lambda_1,\dots,\lambda_M}=
\frac{\prod_j \sinh^{2M}(\lambda_j-\eta/2)\sinh^{2M+1}(\lambda_j+\eta/2)}
{\prod_j \sinh(2\lambda_j) \prod_{j<k} \sinh(\lambda_j-\lambda_k) \sinh(\lambda_j+\lambda_k)}
\times \det L
\end{equation}
with $L$ given by \eqref{benares4}. Similarly, for the overlap with
the dimer and $q$-dimer state we obtain
\begin{equation}
\label{DD}
\begin{split}
 & \skalarszorzat{D}{\lambda_1,\dots,\lambda_M}=
\\
&\hspace{0.5cm}=\frac{\sinh(\eta)^M}{2^{M/2}\cosh(\eta/2)^{M}}
\frac{\prod_j \sinh^{2M}(\lambda_j-\eta/2)\sinh^{2M}(\lambda_j+\eta/2)\cosh(\lambda_j)}
{\prod_j \sinh(2\lambda_j) \prod_{j<k} \sinh(\lambda_j-\lambda_k) \sinh(\lambda_j+\lambda_k)}
\times \det L
\end{split}
\end{equation}
and
\begin{equation}
\begin{split}
\label{qD}
&  \skalarszorzat{qD}{\lambda_1,\dots,\lambda_M}=\\
&\hspace{0.5cm}\frac{\sinh(\eta)^M}{(2\Gamma)^{M/2}}
\frac{\prod_j \sinh^{2M}(\lambda_j-\eta/2)\sinh^{2M}(\lambda_j+\eta/2)e^{\lambda_j}}
{\prod_j \sinh(2\lambda_j) \prod_{j<k} \sinh(\lambda_j-\lambda_k) \sinh(\lambda_j+\lambda_k)}
\times \det L,
\end{split}
\end{equation}
where
\begin{equation*}
\Gamma= 
 \begin{cases}
   \Delta & \text{if } \Delta> 1 \\
   1       & \text{if } \Delta < 1.
  \end{cases}
\end{equation*}
Equations \eqref{N}-\eqref{qD} are the main results of this
work. They are off-shell formulas valid for an arbitrary set of
rapidities provided $\lambda_j+\lambda_k\ne 0$ for any $j,k$. 

The formula \eqref{N} was checked analytically by comparing it to
\eqref{BA3} for $M=1,2,3$ using the computer program
\texttt{Mathematica}. Also, it is easy to see that the relation
\eqref{DN} holds, together with a similar formula for the $q$-dimer state.

\subsection{Pairs of rapidities}

The above determinant formulas are plagued by singularities whenever
$\lambda_j+\lambda_k=0$, in which case further derivatives of
the matrix elements of $L$ need to be taken. Due to the non-linear
nature of the Bethe equations two rapidities with opposite signs only
appear in very special cases. For an even number of roots there are
two types of such states, which will be discussed in the following.

One special configuration is when all rapidities appear in pairs:
\begin{equation}
\{\lambda\}_M=
\{(\lambda_1,-\lambda_1),(\lambda_2,-\lambda_2)\dots(\lambda_{M/2},-\lambda_{M/2})
\}
,\quad \lambda_j\ne 0.
\label{ezisegy}
\end{equation}
The ground states of the XXZ Hamiltonian are of this form for arbitrary $\Delta$.
For such a configuration let $\{\lambda^+\}_{M/2}$ denote the set
$\{\lambda_1,\dots,\lambda_{M/2}\}$. 
For the overlap with the N\'eel state we find
\begin{equation}
\label{N2}
\begin{split}
 & \skalarszorzat{N}{\lambda_1,\dots,\lambda_M}=
(-1)^{\frac{M}{4}(\frac{M}{2}-1)}\times\\ 
&\hspace{2cm}\frac{\prod_j (\sinh(\lambda^+_j-\eta/2)\sinh(\lambda^+_j+\eta/2))^{4M+1}}
{\prod_j \sinh^3(2\lambda_j^+) 
\prod_{j<k} \sinh^4(\lambda^+_j-\lambda^+_k) \sinh^4(\lambda_j^++\lambda^+_k)}
\times \det \tilde L,
\end{split}
\end{equation}
where
\begin{equation}
\label{tl}
\begin{split}
  \tilde L_{jk}
&=q_{2j}(\lambda^+_k),
\qquad k=1\dots M/2\\
 \tilde L_{j,M/2+k}
&=2j(q_{2j+1}(\lambda^+_k)-q_{2j-1}(\lambda^+_k))
,\qquad
k=1\dots M/2
\end{split}
\end{equation}
with $q_a(u)$ given by \eqref{benares4}. Similar formulas can be
written down 
for the overlap with the dimer and $q$-dimer states.

The other special root configuration is when
\begin{equation}
\label{megezisegy}
\{\lambda\}_M=
  \{(\lambda_1,-\lambda_1),(\lambda_2,-\lambda_2)\dots(\lambda_{M/2-1},-\lambda_{M/2-1}),0,i\pi/2 \}.
\end{equation}
Such states appear in the spectrum for both $\Delta>1$ and $\Delta<1$.
In the massive regime $\Delta>1$ regime the first excited state
(which becomes degenerate with the ground state in the infinite volume
limit) is of this form with all rapidities being purely imaginary.
In the $\Delta<1$ regime the rapidity $i\pi/2$ is a so-called  ``negative
parity one-string'' \cite{takahashi-book}, and the remaining
rapidities are either real, or have fixed imaginary part $i\pi/2$, or
they form strings centered on the real axis. 

Let $\{\lambda^+\}_{M/2-1}$ denote the set 
$\{\lambda_1,\dots,\lambda_{M/2-1}\}$. Then the  overlap with the N\'eel
state reads
\begin{equation}
\label{N3}
\begin{split}
 &\skalarszorzat{N}{\lambda_1,\dots,\lambda_M} =(-i)2^{-2M-5} (\sinh(\eta))^{4M+1}
(-1)^{(M/2-1)(M/2-2)/2}\times  \\ 
&\hspace{2cm}\times\frac{\prod_j (\sinh(\lambda^+_j-\eta/2)\sinh(\lambda^+_j+\eta/2))^{4M+1}}
{\prod_j \sinh^7(2\lambda_j^+) 
\prod_{j<k} \sinh^4(\lambda^+_j-\lambda^+_k) \sinh^4(\lambda_j^++\lambda^+_k)}
\times \det \hat L,
\end{split}
\end{equation}
where
\begin{equation}
\label{tl2}
\begin{split}
  \hat L_{jk}
&=q_{2j}(\lambda^+_k),
\qquad k=1\dots M/2-1\\
 \hat L_{j,M/2+k-1}
&=2j(q_{2j+1}(\lambda^+_k)-q_{2j-1}(\lambda^+_k))
,\qquad
k=1\dots M/2-1\\
 \hat L_{j,M-1}&=2j(\coth^{2j+1}(\eta/2)-\coth^{2j-1}(\eta/2))
\\
 \hat L_{j,M}&=2j(\tanh^{2j+1}(\eta/2)-\tanh^{2j-1}(\eta/2)).
\end{split}
\end{equation}
A similar formula holds for the overlap with the $q$-dimer state,
but for the dimer state formula \eqref{DD} gives
\begin{equation}
\label{D0}
  \skalarszorzat{D}{\lambda_1,\dots,\lambda_M}=0.
\end{equation}
This can be understood by noting that
\begin{equation}
\bra{D}B(i\pi/2)=0,
\end{equation}
which can be easily checked by an explicit calculation using the
definition of the $B$-operators.

\section{Numerical results}

In this section we present numerical results for the normalized
overlaps between the three initial states and the ground states of the XXZ
Hamiltonian for different $\Delta$. Let us denote
\begin{equation*}
  S_0^{A}(\Delta,2M)= \frac{|\skalarszorzat{A}{GS}|}
{\sqrt{\skalarszorzat{GS}{GS}}},
\end{equation*}
where $A=N,D,qD$ and $\ket{GS}=\ket{\lambda_1,\dots,\lambda_M}$ is the
ground state of the form \eqref{ezisegy} with
the rapidities being the unique solution of the Bethe equations such
that all of them are purely imaginary (real) for $\Delta>1$
($\Delta<1$), respectively. In the $\Delta>1$ case we also considered
the first excited state which is of the form \eqref{megezisegy}; the
corresponding overlaps will be denoted by $S_1^{A}(\Delta,2M)$.

We solved the Bethe equations for different values of $\Delta$ and $M$ 
and computed the overlaps
together with the norm \eqref{norm}. In order to check the numerical
results we also performed an exact diagonalization of the XXZ
Hamiltonian up to length $2M=16$ and computed the required overlaps. In all cases we found
perfect numerical agreement.

\begin{figure}
  \centering
\psfrag{D}{$\Delta$}
\includegraphics{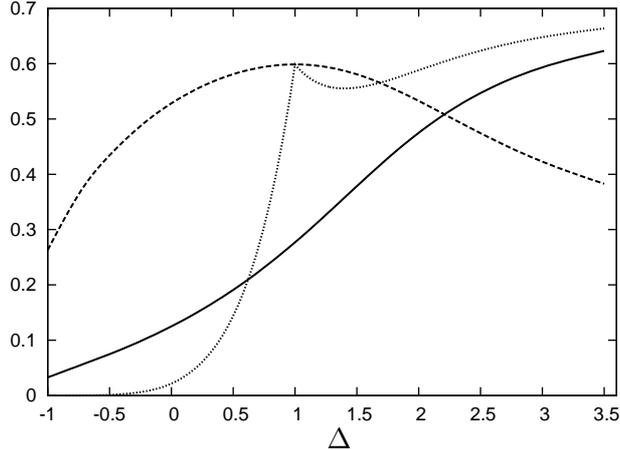}

  \caption{The normalized overlaps with the ground state of the XXZ
    Hamiltonian as a function of $\Delta$ for a spin chain of length
    $2M=12$. The solid line correspond 
    to the N\'eel state, the dashed to the dimer state and the dotted
    to the $q$-dimer state.}
\label{fig1}
\end{figure}

Numerical results for the overlaps in the case $2M=12$ are shown in
figure \ref{fig1}. It can be seen that for large $\Delta$ the q-dimer and
N\'eel states are closer to the ground state, but for smaller $\Delta$
the overlap with the dimer state is bigger. This is expected because
the z-z term in the Hamiltonian tends to ,,freeze'' the
anti-ferromagnetic order, whereas the kinetic x-x and y-y terms make
the positions of the down spins more disperse. The cusp in the overlap
with the $q$-dimer is due to the non-analyticity in its overall
normalization; at $\Delta=1$ the overlap with the dimer and the
$q$-dimer is the same, as expected. Note that the overlap
with the dimer state is largest at the $SU(2)$-symmetric point $\Delta=1$.

The limiting values of the overlaps are as follows. In the large
$\Delta$ limit the ground state turns into the linear combination
\begin{equation*}
  \lim_{\Delta\to\infty}
  \frac{\ket{GS}}{\sqrt{\skalarszorzat{GS}{GS}}}
=\frac{\ket{N}+\ket{AN}}{\sqrt{2}},
\end{equation*}
where $\ket{AN}$ is the anti-N\'eel state. Therefore
\begin{equation*}
\begin{split}
  \lim_{\Delta\to\infty} S_0^{N}(\Delta,2M)&=
\lim_{\Delta\to\infty} S_0^{qD}(\Delta,2M)=\frac{1}{\sqrt{2}}\\
  \lim_{\Delta\to\infty} S_0^{D}(\Delta,2M)&=
(\sqrt{2})^{(1-M)},
\end{split}
\end{equation*}
where we made use of the fact that $M$ is assumed to be even.

The $\Delta\to -1$ limit can be understood by the
well-known transformation
\begin{equation}
\label{UH}
  U^\dagger H_{XXZ}(\Delta=-1) U=
-\sum_{j=1}^{2M}
  (\sigma^x_j\sigma^x_{j+1}+\sigma^y_j\sigma^y_{j+1}+\sigma^z_j\sigma^z_{j+1}-1),
\end{equation}
where
\begin{equation*}
  U=\sum_{j=1}^M \sigma^z_{2j}.
\end{equation*}
The isotropic ferromagnetic Hamiltonian on the r.h.s. of \eqref{UH} has an $2M+1$-fold degenerate
ground state, and the $\Delta\to -1$ limit selects the state with zero
total magnetization. Therefore
\begin{equation*}
  \lim_{\Delta\to -1}\frac{\ket{GS}}{\sqrt{\skalarszorzat{GS}{GS}}}
=U\frac{(S_-)^M}{\sqrt{(2M)!}}
\ket{F_+},
\end{equation*}
where $S_-$ is the global spin lowering operator. This leads to 
\begin{equation*}
\begin{split}
     \lim_{\Delta\to -1} S_0^{N}(\Delta,2M)&=\frac{1}{\sqrt{\binom{2M}{M}}}    \\
   \lim_{\Delta\to -1} S_0^{D}(\Delta,2M)&=\frac{(\sqrt{2})^M}{\sqrt{\binom{2M}{M}}}    \\
   \lim_{\Delta\to -1} S_0^{qD}(\Delta,2M)&= 0.
\end{split}
\end{equation*}
All three results agree with the numerical data obtained from the
determinant formulas.

\begin{table}
  \centering
  \begin{tabular}{|c||c|c|c|c|c|c|}
\hline
  $2M$   & 4 & 8 & 12 & 16 & 20 & 24 \\
\hline
\hline
$\log(S_0(\Delta=1.5))$ & -0.50102 & -0.74718 & -0.97158 & -1.18790&
-1.40000&  -1.60930 \\
\hline
$\log(S_1(\Delta=1.5))$ &  -0.34657 & -0.53730 & -0.75873 & -0.98233&
-1.20450&  -1.42470 \\
\hline
$\log(S_0(\Delta=1.5))$ & 
-0.53850 & -0.88239 & -1.21590 & -1.54680 & -1.87660 & -2.19280 \\
\hline
$\log(S_1(\Delta=1.5))$ &  -0.34657 & -0.59952 & -0.90756 & -1.22790 &
-1.55250 & -1.87900 \\
\hline
$\log(S_0(\Delta=0.5))$ &  -0.61291 &  -1.13460 &  -1.65560&
-2.17700&  -2.69880 & -3.22100 \\
\hline
  \end{tabular}
\caption{Numerical results for the logarithm of the normalized
  overlaps with the N\'eel state. Here $S_0$ denoted the scalar
  product with the ground state, whereas for $S_1$ represents the overlap
  with the first excited state.} 
\label{table1}
\end{table}

We also investigated the $M$-dependence of the overlaps for fixed
$\Delta$; for simplicity here we only considered the case of the
N\'eel state. Numerical results for the logarithm of the scalar
product are given in table \ref{table1} for chain lengths $2M=4\dots
24$. For $\Delta>1$ we also give 
the overlap with the first excited state.

In the large $M$ limit we expect an exponential decay of the overlap:
\begin{equation}
\label{overlapscaling}
  \log(S_0(\Delta,2M))=2M\alpha_0(\Delta) +\beta_0(\Delta)+\dots,\qquad\quad
\alpha_0(\Delta)<0.
\end{equation}
In the recent article \cite{sajat-BQTM} an analytic result was derived
for the linear coefficient $\alpha_0$ using a non-linear integral
equation (NLIE) for the so-called dynamical free energy (the Loschmidt echo
at imaginary times). Moreover it was conjectured that in the
$\Delta>1$ regime the overlap with the first excited state is of the
form \eqref{overlapscaling} with the same linear term: 
\begin{equation}
\label{overlapscaling2}
  \log(S_1(\Delta,2M))=2M\alpha_0(\Delta) +\beta_1(\Delta)+\dots
\end{equation}
Here we attempt to check the exact results and the latter conjecture
of \cite{sajat-BQTM} by comparing them to the finite volume data.

We evaluated the corresponding equations 
 of
\cite{sajat-BQTM}
(eqs. (5.23) and (5.24)) for three values of $\Delta$. Also, we 
performed a linear fit of the data in table \ref{table1}. The
results are shown in table \ref{table2}. 
It can be seen that the
numerical results are close to each other, but the deviation between
the NLIE and finite volume results is in all cases larger than the
error of the linear fit. This discrepancy might be attributed to
additional finite volume effects, for which there is no estimate
available. Numerical data from larger spin chains might give a
better fit; we leave this question for further work.
On the other hand, we note that the two results
agree very well in the massless case $\Delta=0.5$. The derivation 
in \cite{sajat-BQTM} only applies to the massive case, but
 based on this finite volume analysis we conclude that the
 corresponding integral equation is valid also  in the massless regime.

\begin{table}
  \centering
  \begin{tabular}{|c||c|c|}
\hline
&  infinite volume prediction & finite volume fit \\\hline
\hline
\hline
$\alpha_0(\Delta=1.5)$ & -0.05103794 &  -0.0551 $\pm$ 0.0008   \\
\hline
$\alpha_1(\Delta=1.5)$ & -0.05103794 &   -0.0544 $\pm$ 0.0006 \\
\hline
$\alpha_0(\Delta=1.1)$ & -0.08186948 &  -0.0828  $\pm$ 0.0005    \\
\hline
$\alpha_1(\Delta=1.1)$ & -0.08186948 & -0.077      $\pm$    0.002       \\
\hline
$\alpha_0(\Delta=0.5)$ & -0.1308120 &   -0.13039  $\pm$ 0.00002 \\
\hline
  \end{tabular}
\caption{The linear coefficient in the logarithm of the overlap. 
}
\label{table2}
\end{table}

\section{The XXX limit}

\label{sec:xxx}

In this section we consider the special case of the $SU(2)$-symmetric
chain with $\Delta=1$. All relevant formulas for the normalized
overlaps can be obtained by
setting $\eta=i\eps$ and performing the limit
\begin{equation*}
  \eps\to 0, \qquad \frac{\lambda}{\eta}=\text{fixed}.
\end{equation*}
Alternatively the construction of section 3. can be repeated using
the rational R-matrix
\begin{equation}
\tilde   R(u)=
  \begin{pmatrix}
    u+i & & &\\
& u  & i & \\
& i & u & \\
& & & u+i
  \end{pmatrix}.
\label{Rxxx}
\end{equation}
Bethe states are then created as
\begin{equation}
\label{BXXX}
  \ket{\{\lambda\}_n}=\prod_{j=1}^n \tilde B(\lambda_j) \ket{F},
\end{equation}
where $\tilde B(u)$ are the appropriate creation operators obtained
from the monodromy matrix
\begin{equation*}
\tilde   T(u)=\tilde L_{1}(u)\dots \tilde L_{2M}(u),
\end{equation*}
with
\begin{equation*}
  \tilde L_j(u)=\tilde R_{0j}(u-i/2).
\end{equation*}
The Bethe equations become
\begin{equation}
  \label{BAeXXX}
\left(  \frac{\lambda_j-i/2}{\lambda_j+i/2}\right)^{2M}
\prod_{k\ne j}
\frac{\lambda_j-\lambda_k+i}{\lambda_j-\lambda_k-i}=1,
\end{equation}
and the energy is given by
\begin{equation*}
  E=\sum_j e(\lambda_j)\qquad\qquad e(u)=\frac{4}{u^2+1}.
\end{equation*}
The norm of the Bethe states defined in \eqref{BXXX} is
\begin{equation}
\label{normXXX}
  \skalarszorzat{\lambda_1,\dots,\lambda_M}{\lambda_1,\dots,\lambda_M}=
\prod_{j}
(\lambda_j^2+1/4)^{2M}
\prod_{j\ne k} \tilde f(\lambda_j,\lambda_k)\times \det \tilde G,
\end{equation}
where 
\begin{equation*}
\tilde  f(u)=\frac{u+i}{u}
\end{equation*}
and
\begin{equation}
\tilde   G_{jk}=\delta_{j,k}\left(2M \tilde\varphi(\lambda_j,1/2)-\sum_l
    \tilde\varphi(\lambda_j-\lambda_l,1)\right)+
\tilde\varphi(\lambda_j-\lambda_k,1)
\end{equation}
with
\begin{equation}
\tilde  \varphi(a,b)= \frac{2b}{a^2+b^2}.
\end{equation}

The limiting expression for the overlap with the N\'eel state reads
\begin{equation}
\begin{split}
\label{mushroom}  
\skalarszorzat{N}{\lambda_1,\dots,\lambda_M} =
\frac{\prod_j  (\lambda_j-i/2)^{2M}(\lambda_j+i/2)^{2M+1}}
{\prod_j (2\lambda_j) \prod_{j<k} (\lambda_j-\lambda_k) (\lambda_j+\lambda_k)}
\times \det K
\end{split}
\end{equation}
with
\begin{equation}
  K_{jk}=\frac{1}{(\lambda_k-i/2)^{2j}}-\frac{1}{(\lambda_k+i/2)^{2j}}.
\end{equation}
The overlap with the dimer state follows from \eqref{DD} or \eqref{qD}:
\begin{equation}
\begin{split}
\label{mushroomD}  
\skalarszorzat{D}{\lambda_1,\dots,\lambda_M} =
\frac{\prod_j  (\lambda^2_j+1/4)^{2M}}
{2^{M/2}\prod_j (2\lambda_j) \prod_{j<k} (\lambda_j-\lambda_k) (\lambda_j+\lambda_k)}
\times \det K.
\end{split}
\end{equation}
In these formulas we assumed that $\lambda_j\ne\lambda_k$ for any $j\ne k$. In the
XXX case the only special configuration is that of a state
consisting of rapidity pairs only, for which a formula analogous to \eqref{N2} is
easily derived.

\subsection{Overlap with non-highest weight states}

\label{non}

It is known that the Bethe states \eqref{BXXX} are highest weight
states with respect to the global $SU(2)$ symmetry. Non-highest
weight states can be produced by the action of the global spin lowering
operator. It is easy to see from the definition of the $B$-operators
that
\begin{equation*}
  S_- =\frac{1}{i}\lim_{u\to\infty}\frac{1}{ u^{2N-1}} B(u).
\end{equation*}
This relation provides a way to compute overlaps with non-highest weight
states. 

Consider the vector
\begin{equation*}
(S_-)^m \ket{\lambda_{m+1},\lambda_{m+2},\dots,\lambda_M}_{M-m}.
\end{equation*}
It can be obtained by taking an auxiliary Bethe state with rapidities
\begin{equation*}
  (xp_1,xp_2,\dots,xp_m,\lambda_{m+1},\lambda_{m+2},\dots,\lambda_N),
\end{equation*}
and taking the $x\to\infty$ limit. The numbers  $\{p_j\}_m$ can be
chosen arbitrarily with the only requirement that they all be
distinct. 

Substituting these rapidities into \eqref{mushroom} and taking the
$x\to \infty$ limit results in
\begin{equation*}
\begin{split}
&\bra{N}(S_-)^m \ket{\lambda_{m+1},\lambda_{m+2},\dots,\lambda_M}_{M-m}=\\
&\hspace{1.5cm}=\frac{ 2^{m-M}(m!) \prod_{j=m+1}^M
  (\lambda_j-i/2)^{2M}(\lambda_j+i/2)^{2M+1} }
{\prod_{j=m+1}^M \lambda_j \prod_{j<k} (\lambda_j-\lambda_k) (\lambda_j+\lambda_k)}
\times \det \bar K
\end{split}
\end{equation*}
where $\bar K$ is a $(M-m)\times(M-m)$ matrix given by the lower-right
sub-matrix of the original $K$.

In the special case of $m=M$ the overlap is
\begin{equation*}
  \bra{N}(S_-)^M \ket{F}=M!
\end{equation*}
This is the expected result.

Concerning the dimer state the same limiting procedure gives
\begin{equation*}
\begin{split}
\bra{D}(S_-)^m \ket{\lambda_{m+1},\lambda_{m+2},\dots,\lambda_M}_{M-m}=
0,
\end{split}
\end{equation*}
as expected based on the $SU(2)$ invariance of the dimer state.

\section{Conclusions}

In this work we computed determinant formulas for the overlaps between
Bethe states and three special initial states: the N\'eel, the dimer
and the $q$-dimer states. We also considered the special cases when
the Bethe roots form pairs with opposite rapidities. We stress that
our formulas are valid for arbitrary Bethe states, the rapidities do
not need to satisfy the Bethe equations. This makes it possible to
consider the special cases with simple limiting procedures.

It is known that most eigenstates of the XXZ spin chain consist of
strings with different length. Our formulas do not include any
singularities associated with the strings, except when the string
center is exactly at $\lambda=0$. 
In the latter case the formula
\eqref{N3} can be used, whereas for a generic configuration we expect that 
it is sufficient to substitute the approximate positions of the string
rapidities into the determinant, unless the string deviations are large.

It is an important question whether there are more efficient formulas than
those presented here. As it was noted in section \ref{sec:BABA} there
exists an alternative determinant formula which makes use of the Bethe equations;
it can be obtained following the methods of \cite{sajat-Karol}. We did
not include that result here, because it is not convenient for the
states with opposite rapidities and it is not clear if it is better
suited for numerical studies.

An other open question is how to investigate the large volume limit of
the determinant formula, in particular how to extract the leading part
in the logarithm of the overlap.
 First of all, this could lead to an independent
derivation of the exact result of \cite{sajat-BQTM} for the ``overlap
per site'' between the ground state and the different initial states. On the
other hand, this could also help to find the saddle point state of the
``quench-action''  which determines what
configurations are relevant for the long-time behaviour following the
quantum quench \cite{quench-action}. If the string densities of the
saddle point state could be
found then that could lead to an independent check (or possibly an
actual derivation) of the Generalized
Gibbs Ensemble applied to the XXZ chain 
\cite{sajat-xxz-gge,essler-xxz-gge}. 

\vspace{0.4cm} {\bf Note added:}
This article was made available as an arXiv e-print in September 2013,
but it has not been submitted to a journal until now. In the meantime,
two papers by  Brockmann et. al.
\cite{Caux-Neel-overlap1,Caux-Neel-overlap2} presented a more
efficient determinant formula for the overlaps, buiding on the results
of the present paper, and the formula in \cite{sajat-Karol} cited above.

\vspace{1cm}
{\bf Acknowledgments} 

\bigskip

We are grateful to G\'abor Tak\'acs and Jean S\'ebastien Caux for
stimulating discussions. Also, we thank  G\'abor Tak\'acs for useful
comments on the manuscript.

This research was started while the author was employed by the NWO/VENI
grant 016.119.023 at the University of Amsterdam, the Netherlands. 

The second half of the work was realized
in the frames of TAMOP 4.2.4. A/1-11-1-2012-0001
 ,,National Excellence Program -- Elaborating and operating an inland
 student and researcher personal support system convergence program''.
 The project was subsidized 
by the European Union and co-financed by the European Social Fund.

\bigskip

\addcontentsline{toc}{section}{References}
\bibliography{../../../pozsi-general.bib}

\providecommand{\href}[2]{#2}\begingroup\raggedright\begin{thebibliography}{10}

\bibitem{XXX}
H.~Bethe, ``Zur Theorie der Metalle,'' {\em Zeitschrift {f\"ur} Physik} {\bf
  A71} (1931)  205.

\bibitem{sutherland-book}
B.~Sutherland, {\em Beautiful Models}.
\newblock World Scientific Publishing Company, 2004.

\bibitem{takahashi-book}
M.~Takahashi, {\em Thermodynamics of One-Dimensional Solvable Models}.
\newblock Cambridge University Press, 1999.

\bibitem{kluemper-review}
A.~{Kl{\"u}mper}, \href{http://dx.doi.org/10.1007/BFb0119598}{``{Integrability
  of Quantum Chains: Theory and Applications to the Spin-1/2 XXZ Chain},''} in
  {\em Quantum Magnetism}, {U.~Schollw{\"o}ck, J.~Richter, D.~J.~J.~Farnell, \&
  R.~F.~Bishop }, ed., vol.~645 of {\em Lecture Notes in Physics, Berlin
  Springer Verlag}, p.~349.
\newblock 2004.
\newblock \href{http://arxiv.org/abs/arXiv:cond-mat/0502431}{{\tt
  arXiv:cond-mat/0502431}}.

\bibitem{slavnov-overlaps}
N.~A. Slavnov, ``Calculation of scalar products of wave functions and form
  factors in the framework of the algebraic Bethe ansatz,''
  \href{http://dx.doi.org/10.1007/BF01016531}{{\em Theoretical and Mathematical
  Physics} {\bf 79} (1989)  502--508}.

\bibitem{maillet-inverse-scatt}
N.~{Kitanine}, J.~M. {Maillet}, and V.~{Terras}, ``{Form factors of the XXZ
  Heisenberg spin-1/2 finite chain},''
  \href{http://dx.doi.org/10.1016/S0550-3213(99)00295-3}{{\em Nucl. Phys. B}
  {\bf 554} (1999)  647--678},
  \href{http://arxiv.org/abs/arXiv:math-ph/9807020}{{\tt
  arXiv:math-ph/9807020}}.

\bibitem{goehmann-korepin-inverse}
F.~G\"ohmann and V.~E. Korepin, ``Solution of the quantum inverse problem,''
  \href{http://dx.doi.org/10.1088/0305-4470/33/6/308}{{\em J. Phys. A} {\bf 33}
  (2000) no.~6, 1199}, \href{http://arxiv.org/abs/hep-th/9910253}{{\tt
  arXiv:hep-th/9910253}}.

\bibitem{KitanineMailletTerras-XXZ-corr1}
N.~Kitanine, J.~M. Maillet, and V.~Terras, ``Correlation functions of the XXZ
  Heisenberg spin-1/2 chain in a magnetic field,'' \href{http://dx.doi.org/DOI:
  10.1016/S0550-3213(99)00619-7}{{\em Nucl. Phys. B} {\bf 567} (2000) no.~3,
  554 -- 582}, \href{http://arxiv.org/abs/math-ph/9907019}{{\tt
  arXiv:math-ph/9907019}}.

\bibitem{spin-spin-XXZ}
N.~Kitanine, J.~M. Maillet, N.~A. Slavnov, and V.~Terras, ``{Spin-spin
  correlation functions of the XXZ-1/2 Heisenberg chain in a magnetic field},''
  \href{http://dx.doi.org/10.1016/S0550-3213(02)00583-7}{{\em Nucl. Phys.} {\bf
  B641} (2002)  487--518}, \href{http://arxiv.org/abs/hep-th/0201045}{{\tt
  arXiv:hep-th/0201045}}.

\bibitem{QTM1}
F.~{G{\"o}hmann}, A.~{Kl{\"u}mper}, and A.~{Seel}, ``{Integral representations
  for correlation functions of the XXZ chain at finite temperature},''
  \href{http://dx.doi.org/10.1088/0305-4470/37/31/001}{{\em Journal of Physics
  A Mathematical General} {\bf 37} (2004)  7625--7651},
  \href{http://arxiv.org/abs/arXiv:hep-th/0405089}{{\tt arXiv:hep-th/0405089}}.

\bibitem{XXZ-factorization-recent-osszefoglalo}
J.~{Sato}, B.~{Aufgebauer}, H.~{Boos}, F.~{G{\"o}hmann}, A.~{Kl{\"u}mper},
  M.~{Takahashi}, and C.~{Trippe}, ``{Computation of Static Heisenberg-Chain
  Correlators: Control over Length and Temperature Dependence},''
  \href{http://dx.doi.org/10.1103/PhysRevLett.106.257201}{{\em Physical Review
  Letters} {\bf 106} (2011) no.~25, 257201},
  \href{http://arxiv.org/abs/1105.4447}{{\tt arXiv:1105.4447
  [cond-mat.str-el]}}.

\bibitem{Silva-quench-colloquium}
A.~Polkovnikov, K.~Sengupta, A.~Silva, and M.~Vengalattore,
  ``\textit{Colloquium} : Nonequilibrium dynamics of closed interacting quantum
  systems,'' \href{http://dx.doi.org/10.1103/RevModPhys.83.863}{{\em Rev. Mod.
  Phys.} {\bf 83} (2011)  863--883}.

\bibitem{2013PhRvA..87e3628I}
D.~{Iyer}, H.~{Guan}, and N.~{Andrei}, ``{Exact formalism for the quench
  dynamics of integrable models},''
  \href{http://dx.doi.org/10.1103/PhysRevA.87.053628}{{\em Physical Review A}
  {\bf 87} (2013) no.~5, 053628}, \href{http://arxiv.org/abs/1304.0506}{{\tt
  arXiv:1304.0506 [cond-mat.quant-gas]}}.

\bibitem{marci-ll-quench1}
M.~{Kormos}, A.~{Shashi}, Y.-Z. {Chou}, J.-S. {Caux}, and A.~{Imambekov},
  ``{Interaction quenches in the 1D Bose gas},'' {\em Phys. Rev. B} {\bf 88}
  (2013)  205131, \href{http://arxiv.org/abs/1305.7202}{{\tt arXiv:1305.7202
  [cond-mat.stat-mech]}}.

\bibitem{spyros2}
M.~{Collura}, S.~{Sotiriadis}, and P.~{Calabrese}, ``{Quench dynamics of a
  Tonks-Girardeau gas released from a harmonic trap},'' {\em Journal of
  Statistical Mechanics: Theory and Experiment} {\bf 2013} (2013) no.~09,
  P09025, \href{http://arxiv.org/abs/1306.5604}{{\tt arXiv:1306.5604
  [cond-mat.quant-gas]}}.

\bibitem{marci-ll-quench2}
M.~{Kormos}, M.~{Collura}, and P.~{Calabrese}, ``{Analytic results for a
  quantum quench from free to hard-core one dimensional bosons},''
  \href{http://dx.doi.org/10.1103/PhysRevA.89.013609}{{\em Phys. Rev. A} {\bf
  89} (2014)  013609}, \href{http://arxiv.org/abs/1307.2142}{{\tt
  arXiv:1307.2142 [cond-mat.quant-gas]}}.

\bibitem{caux-stb-LL-BEC-quench}
J.~De~Nardis, B.~Wouters, M.~Brockmann, and J.-S. Caux, ``Solution for an
  interaction quench in the Lieb-Liniger Bose gas,''
  \href{http://dx.doi.org/10.1103/PhysRevA.89.033601}{{\em Phys. Rev. A} {\bf
  89} (2014)  033601}, \href{http://arxiv.org/abs/1308.4310}{{\tt
  arXiv:1308.4310 [cond-mat.stat-mech]}}.

\bibitem{sajat-xxz-gge}
B.~Pozsgay, ``The generalized Gibbs ensemble for Heisenberg spin chains,'' {\em
  Journal of Statistical Mechanics: Theory and Experiment} {\bf 2013} (2013)
  no.~07, 3, \href{http://arxiv.org/abs/1304.5374}{{\tt arXiv:1304.5374
  [cond-mat.stat-mech]}}.

\bibitem{essler-xxz-gge}
M.~{Fagotti} and F.~H.~L. {Essler}, ``{Stationary behaviour of observables
  after a quantum quench in the spin-1/2 Heisenberg XXZ chain},''
  \href{http://dx.doi.org/10.1088/1742-5468/2013/07/P07012}{{\em Journal of
  Statistical Mechanics: Theory and Experiment} {\bf 7} (2013)  12},
  \href{http://arxiv.org/abs/1305.0468}{{\tt arXiv:1305.0468
  [cond-mat.stat-mech]}}.

\bibitem{rigol-gge}
M.~Rigol, V.~Dunjko, V.~Yurovsky, and M.~Olshanii, ``Relaxation in a Completely
  Integrable Many-Body Quantum System: An Ab Initio Study of the Dynamics of
  the Highly Excited States of 1D Lattice Hard-Core Bosons,''
  \href{http://dx.doi.org/10.1103/PhysRevLett.98.050405}{{\em Phys. Rev. Lett.}
  {\bf 98} (2007) no.~5, 050405},
  \href{http://arxiv.org/abs/arXiv:cond-mat/0604476}{{\tt
  arXiv:cond-mat/0604476}}.

\bibitem{ising-quench-1}
P.~{Calabrese}, F.~H.~L. {Essler}, and M.~{Fagotti}, ``{Quantum quench in the
  transverse field Ising chain: I. Time evolution of order parameter
  correlators},''
  \href{http://dx.doi.org/10.1088/1742-5468/2012/07/P07016}{{\em Journal of
  Statistical Mechanics: Theory and Experiment} {\bf 7} (2012)  16},
  \href{http://arxiv.org/abs/1204.3911}{{\tt arXiv:1204.3911
  [cond-mat.quant-gas]}}.

\bibitem{ising-quench-2}
P.~{Calabrese}, F.~H.~L. {Essler}, and M.~{Fagotti}, ``{Quantum quenches in the
  transverse field Ising chain: II. Stationary state properties},''
  \href{http://dx.doi.org/10.1088/1742-5468/2012/07/P07022}{{\em Journal of
  Statistical Mechanics: Theory and Experiment} {\bf 7} (2012)  22},
  \href{http://arxiv.org/abs/1205.2211}{{\tt arXiv:1205.2211
  [cond-mat.stat-mech]}}.

\bibitem{sajat-BQTM}
B.~{Pozsgay}, ``{Dynamical free energy and the Loschmidt-echo for a class of
  quantum quenches in the Heisenberg spin chain},'' {\em Journal of Statistical
  Mechanics: Theory and Experiment} {\bf 2013} (2013) no.~10, P10028,
  \href{http://arxiv.org/abs/1308.3087}{{\tt arXiv:1308.3087
  [cond-mat.stat-mech]}}.

\bibitem{Fagotti-Loschmidt}
M.~{Fagotti}, ``{Dynamical Phase Transitions as Properties of the Stationary
  State: Analytic Results after Quantum Quenches in the Spin-1/2 XXZ Chain},''
  {\em ArXiv e-prints} (2013)  , \href{http://arxiv.org/abs/1308.0277}{{\tt
  arXiv:1308.0277 [cond-mat.stat-mech]}}.

\bibitem{quench-action}
J.-S. {Caux} and F.~H.~L. {Essler}, ``{Time Evolution of Local Observables
  After Quenching to an Integrable Model},''
  \href{http://dx.doi.org/10.1103/PhysRevLett.110.257203}{{\em Physical Review
  Letters} {\bf 110} (2013) no.~25, 257203},
  \href{http://arxiv.org/abs/1301.3806}{{\tt arXiv:1301.3806
  [cond-mat.stat-mech]}}.

\bibitem{XXZ1}
R.~Orbach, ``Linear Antiferromagnetic Chain with Anisotropic Coupling,''
  \href{http://dx.doi.org/10.1103/PhysRev.112.309}{{\em Phys. Rev.} {\bf 112}
  (1958) no.~2, 309--316}.

\bibitem{XXZ2}
L.~R. Walker, ``Antiferromagnetic Linear Chain,''
  \href{http://dx.doi.org/10.1103/PhysRev.116.1089}{{\em Phys. Rev.} {\bf 116}
  (1959) no.~5, 1089--1090}.

\bibitem{XXZ3}
C.~N. Yang and C.~P. Yang, ``One-Dimensional Chain of Anisotropic Spin-Spin
  Interactions. I. Proof of Bethe's Hypothesis for Ground State in a Finite
  System,'' \href{http://dx.doi.org/10.1103/PhysRev.150.321}{{\em Phys. Rev.}
  {\bf 150} (1966) no.~1, 321--327}.

\bibitem{MG}
C.~K. Majumdar and D.~K. Ghosh, ``On Next-Nearest-Neighbor Interaction in
  Linear Chain. I,'' \href{http://dx.doi.org/10.1063/1.1664978}{{\em Journal of
  Mathematical Physics} {\bf 10} (1969) no.~8, 1388--1398}.

\bibitem{q-dimer}
M.~T. {Batchelor} and C.~M. {Yung}, ``{q-Deformations of Quantum Spin Chains
  with Exact Valence-Bond Ground States},''
  \href{http://dx.doi.org/10.1142/S021797929400155X}{{\em International Journal
  of Modern Physics B} {\bf 8} (1994)  3645--3654},
  \href{http://arxiv.org/abs/arXiv:cond-mat/9403080}{{\tt
  arXiv:cond-mat/9403080}}.

\bibitem{gaudin-LL-norms}
M.~Gaudin, ``La function d'onde de Bethe pour les mod\`eles exacts de la
  m\'ecanique statistique,'' {\em Commisariat \'a l'\'energie atomique, Paris}
  (1983)  .

\bibitem{korepin-norms}
V.~E. Korepin, ``Calculation of norms of Bethe wave functions,''
  \href{http://dx.doi.org/10.1007/BF01212176}{{\em Comm. Math. Phys.} {\bf 86}
  (1982)  391}.

\bibitem{sklyanin-boundary}
E.~K. Sklyanin, ``Boundary conditions for integrable quantum systems,''
  \href{http://dx.doi.org/10.1088/0305-4470/21/10/015}{{\em J. Phys. A} {\bf
  21} (1988) no.~10, 2375}.

\bibitem{openXXZ1}
N.~{Kitanine}, K.~K. {Kozlowski}, J.~M. {Maillet}, G.~{Niccoli}, N.~A.
  {Slavnov}, and V.~{Terras}, ``{Correlation functions of the open XXZ chain:
  I},'' \href{http://dx.doi.org/10.1088/1742-5468/2007/10/P10009}{{\em J. Stat.
  Mech.} {\bf 10} (2007)  9}, \href{http://arxiv.org/abs/0707.1995}{{\tt
  arXiv:0707.1995 [hep-th]}}.

\bibitem{tsuchiya-det}
O.~Tsuchiya, ``Determinant formula for the six-vertex model with reflecting
  end,'' \href{http://dx.doi.org/10.1063/1.532606}{{\em Journal of Mathematical
  Physics} {\bf 39} (1998) no.~11, 5946--5951},
  \href{http://arxiv.org/abs/solv-int/9804010}{{\tt arXiv:solv-int/9804010}}.

\bibitem{sajat-Karol}
K.~K. Kozlowski and B.~Pozsgay, ``Surface free energy of the open XXZ spin-1/2
  chain,'' \href{http://dx.doi.org/10.1088/1742-5468/2012/05/P05021}{{\em J.
  Stat. Mech.} {\bf 2012} (2012) no.~05, 21}.

\bibitem{six-vertex-partition-function}
A.~G. Izergin, D.~A. Coker, and V.~E. Korepin, ``Determinant formula for the
  six-vertex model,'' \href{http://dx.doi.org/10.1088/0305-4470/25/16/010}{{\em
  J. Phys. A} {\bf 25} (1992) no.~16, 4315}.

\bibitem{Caux-Neel-overlap1}
M.~{Brockmann}, J.~{De Nardis}, B.~{Wouters}, and J.-S. {Caux}, ``{A
  Gaudin-like determinant for overlaps of N{\'e}el and XXZ Bethe states},''
  {\em Journal of Physics A: Mathematical and Theoretical} {\bf 47} (2014)
  no.~14, 145003, \href{http://arxiv.org/abs/1401.2877}{{\tt arXiv:1401.2877}}.

\bibitem{Caux-Neel-overlap2}
M.~{Brockmann}, J.~{De Nardis}, B.~{Wouters}, and J.-S. {Caux}, ``{N{\'e}el-XXZ
  state overlaps: odd particle numbers and Lieb-Liniger scaling limit},'' {\em
  ArXiv e-prints} (2014)  , \href{http://arxiv.org/abs/1403.7469}{{\tt
  arXiv:1403.7469 [cond-mat.stat-mech]}}.

\end{thebibliography}\endgroup
\bibliographystyle{utphys}

\end{document}